\documentclass[12pt]{article}

\usepackage{amsmath,amssymb,amsbsy,amsfonts,amsthm,latexsym,
                        amsopn,amstext,amsxtra,euscript,amscd,color}
\usepackage{color,xcolor}
\usepackage{latexsym}
\usepackage{cite}
\usepackage{amsfonts}
\usepackage{amsfonts,mathrsfs}
\usepackage{graphicx}
\usepackage{psfrag}
\usepackage{subfigure}
\usepackage{url}
\usepackage{stfloats}
\usepackage{amsmath}

\newtheorem{theorem}{Theorem}
\newtheorem{lemma}{Lemma}
\newtheorem{corollary}{Corollary}

\newcommand{\quash}[1]{}

\begin{document}

\title{On $k$-error linear complexity of pseudorandom binary sequences derived from Euler quotients}
\author{Zhixiong Chen$^{1}$, Vladimir Edemskiy$^{2}$, Pinhui Ke$^{3}$, Chenhuang Wu$^{1}$\\
1. Provincial Key Laboratory of Applied Mathematics,\\ Putian University, Putian, Fujian 351100, P.R. China\\
2. Department of Applied Mathematics and Informatics,\\ Novgorod State University, Veliky Novgorod, 173003, Russia\\
3. Fujian Provincial Key Laboratory of Network Security and Cryptology,\\
College of Mathematics and Informatics,\\
Fujian Normal University,
Fuzhou, Fujian, 350117, P. R. China}

%\date{}
\maketitle

\def\F{\mathbb{F}}
\def\Z{\mathbb{Z}}

\begin{abstract}
We investigate the $k$-error linear complexity of pseudorandom
binary sequences of period $p^{\mathfrak{r}}$ 
derived from the Euler quotients modulo $p^{\mathfrak{r}-1}$, a power of an odd prime $p$ for $\mathfrak{r}\geq 2$. 
When $\mathfrak{r}=2$, this is just the case of polynomial quotients (including Fermat quotients) modulo $p$, which has been studied in 
an earlier work of Chen, Niu and Wu. In this work, 
we establish a recursive relation
on the $k$-error linear complexity of the sequences for the case of $\mathfrak{r}\geq 3$.
We also state the exact values of the $k$-error
linear complexity for the case of $\mathfrak{r}=3$.
From the results, we can find that the $k$-error linear complexity of the sequences (of period $p^{\mathfrak{r}}$) does not decrease dramatically for $k<p^{\mathfrak{r}-2}(p-1)^2/2$.  
\end{abstract}



\section{Introduction}
For an odd prime $p$,  integers $r\geq 1$ and  $u$ with
$\gcd(u,p)=1$, the {\it Euler quotient\/} modulo $p^r$, denoted by $Q_{r}(u)$,
is defined as the unique integer belonging to $\{0,1,\ldots,p^r-1\}$ by
$$
Q_{r}(u) \equiv \frac{u^{\varphi(p^r)} -1}{p^r} \pmod {p^r},
$$
where $\varphi(-)$ is the Euler totient function with
$\varphi(p^r)=p^{r-1}(p-1)$, see, e.g., \cite{ADS,Sha} for details.
In addition, we define $Q_{r}(u) = 0$ if $p|u$. In particular,
$Q_{1}(u)$ is called the {\it Fermat quotient}. Many number
theoretic problems have been studied for Fermat and Euler quotients
in \cite{ADS,BFKS,Chang,CW2,CW3,OS,Sha,S,S2010,S2011,S2011b,SW} and
references therein. In the past decade, Fermat and Euler quotients
have also been studied from the viewpoint of cryptography, families
of pseudorandom sequences with  nice cryptographic features are
derived from  Fermat and Euler quotients, see
\cite{C,CD,CDR2015,CNW,COW,DCH,GW,NCD,OS}.

In this correspondence, we still concentrate on a family of binary sequences defined by Euler quotients \cite{CDR2015,DCH}.
For a fixed $\mathfrak{r}\ge 2$,
$(s_n^{(\mathfrak{r})})$ is defined as
\begin{equation}\label{binarythreshold}
s_n^{(\mathfrak{r})}=\left\{
\begin{array}{ll}
0, & \mathrm{if}\,\ 0\leq Q_{\mathfrak{r}-1}(n)/p^{\mathfrak{r}-1}< \frac{1}{2},\\
1, & \mathrm{if}\,\ \frac{1}{2}\leq Q_{\mathfrak{r}-1}(n)/p^{\mathfrak{r}-1}< 1,
\end{array}
\right. ~~~ n\ge 0.
\end{equation}
We always use $\mathfrak{r}$ as a fixed parameter for the sequences and $r$ as a general variable.

Due to the fact \cite{ADS} that
\begin{equation}\label{eq:add struct2}
Q_{r}(u+kp^{r})\equiv Q_{r}(u) - kp^{r-1}u^{-1} \pmod {p^r}, ~ \gcd(u,p) = 1, ~ k\in\mathbb{Z},
\end{equation}
for $r\geq 1$, we see that $(s_n^{(\mathfrak{r})})$ is $p^{\mathfrak{r}}$-periodic.

The linear complexity and the $k$-error linear complexity of $(s_n^{(2)})$, i.e. $\mathfrak{r}=2$,  have been investigated in \cite{CD} and in \cite{CNW},
respectively. In fact, in \cite{CNW} the $k$-error linear complexity has been considered for a general quotient called polynomial quotient.
The trace representation of  $(s_n^{(2)})$ has been presented in \cite{C}.
For $\mathfrak{r}>2$, the linear complexity of $(s_n^{(\mathfrak{r})})$ has been investigated in \cite{DCH} and the trace representation of  $(s_n^{(\mathfrak{r})})$ has been given in \cite{CDR2015}. While the $k$-error linear complexity of $(s_n^{(\mathfrak{r})})$ for $\mathfrak{r}>2$ is still open. This leads to the work in the correspondence.

We organize this correspondence as follows. In Section 2, we firstly introduce some necessary lemmas and an important technique for the proof.
 Then we  prove the main theorem for $\mathfrak{r}\ge 3$. In Section 3, we draw a conclusion and give some further problems.

We conclude this section by recalling the notions of the linear
complexity and the $k$-error linear complexity. Let $\mathbb{F}$ be a field.  For a $T$-periodic
sequence $(h_n)$ over $\mathbb{F}$, we recall that the
\emph{linear complexity} over $\mathbb{F}$, denoted by  $LC^{\mathbb{F}}((h_n))$, is the least order $L$  such that $(h_n)$ satisfies
$$
h_{n+L} = c_{L-1}h_{n+L-1} +\ldots +c_1h_{n+1}+ c_0h_n\quad
\mathrm{for}\,\ n \geq 0,
$$
where $c_0\neq 0, c_1, \ldots,
c_{L-1}\in \mathbb{F}$.
Let
$$
H(X)=h_0+h_1X+h_2X^2+\ldots+h_{T-1}X^{T-1}\in \mathbb{F}[X],
$$
which is called the \emph{generating polynomial} of $(h_n)$. Then the linear
complexity over $\mathbb{F}$ of $(h_n)$ is computed by
$$
  LC^{\mathbb{F}}((h_n)) =T-\deg\left(\mathrm{gcd}(X^T-1,
  ~H(X))\right),
$$
see, e.g. \cite{CDR} for details. For integers $k\ge 0$, the \emph{$k$-error linear complexity} over $\mathbb{F}$ of $(h_n)$, denoted by $LC^{\mathbb{F}}_k(h_n))$, is the smallest linear complexity (over $\mathbb{F}$) that can be
obtained by changing at most $k$ terms of the sequence per period, see \cite{SM}, and see \cite{DXS} for the related even earlier defined \emph{sphere complexity}.  Clearly $LC^{\mathbb{F}}_0((h_n))=LC^{\mathbb{F}}((h_n))$ and
$$
T\ge LC^{\mathbb{F}}_0((h_n))\ge LC^{\mathbb{F}}_1((h_n))\ge \ldots \ge LC^{\mathbb{F}}_l((h_n))=0,
$$
where $l$ equals the number of nonzero terms of $(h_n)$ per period, i.e., the weight of $(h_n)$.

The linear complexity and the $k$-error linear complexity are
important cryptographic characteristics of sequences and provide
information on the predictability and thus unsuitability for
cryptography. For a sequence to be cryptographically strong, its
linear complexity should be large, and at the same time not
significantly reduced by changing a few terms.

\section{ $k$-Error linear complexity}

We denote by
$\mathbb{Z}_{m}=\{0,1,\ldots, m-1\}$ the residue class ring
modulo $m$ and by $\mathbb{Z}_{m}^*$ the unit group of
$\mathbb{Z}_{m}$. For $r\geq 1$, by (\ref{eq:add struct2}) and
$$
Q_{r}(uv)=Q_{r}(u)+Q_{r}(v) \pmod {p^r},~\gcd(uv,p)=1,
$$
the quotient $Q_r(-)$ defines a group epimorphism from $\mathbb{Z}^*_{p^{r+1}}$ to $\mathbb{Z}_{p^{r}}$.
Let
$$
D_l^{(p^{r+1})}=\{u: 0\le u< p^{r+1},~ \gcd(u,p)=1,~ Q_{r}(u)=l\}
$$
for $l=0,1,\ldots,p^r-1$. Clearly, $D_0^{(p^{r+1})}, D_1^{(p^{r+1})},\ldots,
D_{p^r-1}^{(p^{r+1})}$ form a partition of $\mathbb{Z}_{p^{r+1}}^*$.

Since $\mathbb{Z}^*_{p^{r+1}}$ is cyclic, we choose the element $g\in\mathbb{Z}^*_{p^{r+1}}$ as a
 generator ($g$ is also called a primitive  element of $\mathbb{Z}^*_{p^{r+1}}$).
 We remark that one can  always  choose a primitive element $g$ such that $Q_{r}(g)=1$, see a short proof in \cite{CDR2015}.
Then $D_0^{(p^{r+1})}, D_1^{(p^{r+1})},\ldots,
D_{p^r-1}^{(p^{r+1})}$  defined above can be obtained in a following way,
$$
D_l^{(p^{r+1})}=\{g^{l+kp^r} \pmod {p^{r+1}}: ~0 \le k < p-1 \}
$$
for $0\le l< p^r$. Then each $|D_l^{(p^{r+1})}|=p-1$, here and hereafter $|Z|$ means the cardinality of a set $Z$.

Let $\mathfrak{r}\geq 2$ and
$$
\mathcal{C}_0=\bigcup_{l=0}^{(p^{\mathfrak{r}-1}-1)/2}D_{l}^{(p^{\mathfrak{r}})}~~~\bigcup p\mathbb{Z}_{p^{\mathfrak{r}-1}}, ~~~~
\mathcal{C}_1=\bigcup_{l=(p^{\mathfrak{r}-1}+1)/2}^{p^{\mathfrak{r}-1}-1}D_{l}^{(p^{\mathfrak{r}})}
$$
then
one can define
$(s_n^{(\mathfrak{r})})$ equivalently by
\begin{equation}\label{euler-binary}
s_n^{(\mathfrak{r})}=\left\{
\begin{array}{ll}
0, & \mathrm{if} ~ n\pmod {p^{\mathfrak{r}}} \in \mathcal{C}_0,\\
1, &  \mathrm{if} ~  n\pmod {p^{\mathfrak{r}}} \in \mathcal{C}_1.
\end{array}
\right.
\end{equation}

The linear complexity of $(s_n^{(\mathfrak{r})})$ has been investigated
in \cite{CD} for $\mathfrak{r}=2$ and in \cite{DCH} for $\mathfrak{r}>2$, respectively,  that is

\begin{theorem}\label{LC}(\cite{CD,DCH})
Let $(s_n^{(\mathfrak{r})})$ be the binary sequence over
$\mathbb{F}_2$ defined in Eq.(\ref{binarythreshold}) or
Eq.(\ref{euler-binary}). If $2^{p-1}\not\equiv 1 \pmod {p^2}$, then
the linear complexity  of   $(s_n^{(\mathfrak{r})})$  for $\mathfrak{r}\geq 2$ satisfies
\[
LC^{\F_2}((s_n^{(\mathfrak{r})}))=\left\{
\begin{array}{cl}
p^{\mathfrak{r}}-p, & \mathrm{if}\,\ p \equiv 1 \pmod 4, \\
p^{\mathfrak{r}}-p, & \mathrm{if}\,\ p \equiv 3 \pmod 4 ~~\mathrm{and ~~odd}~~\mathfrak{r}, \\
p^{\mathfrak{r}}-1, & \mathrm{if}\,\ p \equiv 3 \pmod 4 ~~\mathrm{and ~~even}~~\mathfrak{r}.
\end{array}
\right.\\
\]
\end{theorem}

 The $k$-error
linear complexity of $(s_n^{(\mathfrak{r})})$ has been investigated
in \cite{CNW} for $\mathfrak{r}=2$, that is

\begin{theorem}\label{klc-2-primitive}(\cite[Corollary 1]{CNW})
Let $(s_n^{(\mathfrak{r})})$ be the binary sequence over
$\mathbb{F}_2$ defined in Eq.(\ref{binarythreshold}) or
Eq.(\ref{euler-binary}).
 If $2$ is a primitive root modulo $p^2$, then
the $k$-error linear complexity  of  $(s_n^{(\mathfrak{r})})$ for $\mathfrak{r}=2$ satisfies
\[
LC^{\F_2}_k((s_n^{(2)}))=\left\{
\begin{array}{cl}
p^2-p, & \mathrm{if}\,\ 0\le k<(p-1)^2/2, \\
0, & \mathrm{if}\,\ k\ge (p-1)^2/2,
\end{array}
\right.\\
\]
 if $p \equiv 1 \pmod 4$, and otherwise
\[
 LC^{\F_2}_k((s_n^{(2)}))=\left\{
\begin{array}{cl}
p^2-1, & \mathrm{if}\,\ k=0, \\
p^2-p+1, & \mathrm{if}\,\ 1\le k<p-1, \\
p^2-p, & \mathrm{if}\,\ p-1\le k< (p-1)^2/2, \\
0, & \mathrm{if}\,\ k\ge (p-1)^2/2.
\end{array}
\right.
\]
\end{theorem}

Our main goal is to extend Theorem \ref{klc-2-primitive} to the case $\mathfrak{r}>2$, which is described in the following theorem.
 For our purpose,
we define another binary sequence
$(\overline{s}_n^{(\mathfrak{r})})$ by
\begin{equation}\label{euler-binary-complement}
\overline{s}_n^{(\mathfrak{r})}=\left\{
\begin{array}{ll}
1, & \mathrm{if} ~ n\pmod {p^{\mathfrak{r}}} \in \bigcup_{l=0}^{(p^{\mathfrak{r}-1}-1)/2}D_{l}^{(p^{\mathfrak{r}})},\\
0, &  \mathrm{otherwise}.
\end{array}
\right.
\end{equation}

\begin{theorem}\label{klc-r}(Main theorem).
Let $(s_n^{(\mathfrak{r})})$ be the binary sequences over $\mathbb{F}_2$ defined in
Eq.(\ref{binarythreshold}) or Eq.(\ref{euler-binary}).
 If $2$ is a primitive root modulo $p^2$, then
the $k$-error linear complexity  of
$(s_n^{(\mathfrak{\mathfrak{r}})})$ for $\mathfrak{r}\ge 3$ satisfies
\[
 LC^{\F_2}_k((s_n^{(\mathfrak{r})}))=\left\{
\begin{array}{cl}
p^{\mathfrak{r}}-p^{\mathfrak{r}-1}+LC^{\F_2}_k((s_n^{(\mathfrak{r}-1)})), & \mathrm{if}\,\ 0\leq k<(p^{\mathfrak{r}-2}-1)(p-1)/2,\\
p^{\mathfrak{r}}-p^{\mathfrak{r}-1}, & \mathrm{if}\,\ (p^{\mathfrak{r}-2}-1)(p-1)/2 \leq k<p^{\mathfrak{r}-2}(p-1)^2/2,\\
p^{\mathfrak{r}-1}-p, & \mathrm{if}\,\ k=p^{\mathfrak{r}-2}(p-1)^2/2,
\end{array}
\right.
\]
 if $p \equiv 1 \pmod 4$, and otherwise
\[
 LC^{\F_2}_k((s_n^{(\mathfrak{r})}))=\left\{
\begin{array}{cl}
p^{\mathfrak{r}}-p^{\mathfrak{r}-1}+LC^{\F_2}_k((\overline{s}_n^{(\mathfrak{r}-1)})), & \mathrm{if}\,\ 0\leq k<(p^{\mathfrak{r}-2}+1)(p-1)/2,\\
p^{\mathfrak{r}}-p^{\mathfrak{r}-1}, & \mathrm{if}\,\ (p^{\mathfrak{r}-2}+1)(p-1)/2 \leq k<p^{\mathfrak{r}-2}(p-1)^2/2,\\
p^{\mathfrak{r}-1}-p, & \mathrm{if}\,\ k=p^{\mathfrak{r}-2}(p-1)^2/2\,\ \mathrm{and} \,\ \mathfrak{r} \,\  \mathrm{is\,\ even}, \\
p^{\mathfrak{r}-1}-1, & \mathrm{if}\,\ k=p^{\mathfrak{r}-2}(p-1)^2/2 \,\ \mathrm{and} \,\ \mathfrak{r} \,\  \mathrm{is\,\ odd},
\end{array}
\right.
\]
where $(\overline{s}_n^{(\mathfrak{r})})$ is defined in Eq.(\ref{euler-binary-complement}).
\end{theorem}

Below we make some preparations for the proof of the main result.

\subsection{Auxiliary Lemmas}

Throughout this work, we use
$$
d_{l}^{(p^r)}(X)=\sum\limits_{n\in D_{l}^{(p^r)}}X^{n} \in
\mathbb{F}_2[X], ~~~ 0\leq l<p^{r-1}
$$
for $r\geq 2$. Write
\begin{equation}\label{gene-poly}
S^{(p^{r})}(X)=\sum\limits_{l=(p^{r-1}+1)/2}^{p^{r-1}-1}d_{l}^{(p^{r})}(X)
~~~ \mathrm{and}~~~
\overline{S}^{(p^{r})}(X)=\sum\limits_{l=0}^{(p^{r-1}-1)/2}d_{l}^{(p^{r})}(X).
\end{equation}
We see that $S^{(p^{\mathfrak{r}})}(X)$ is the generating polynomial of $(s_n^{(\mathfrak{r})})$, and
$\overline{S}^{(p^{\mathfrak{r}})}(X)$ is of $(\overline{s}_n^{(\mathfrak{r})})$.

It is straightforward to verify  that
\begin{equation}\label{SX}
S^{(p^{r})}(X)=
\sum\limits_{i=(p+1)/2}^{p-1}~~\sum\limits_{j=0}^{(p^{r-2}-1)/2}d_{ip^{r-2}+j}^{(p^{r})}(X)+
\sum\limits_{i=(p-1)/2}^{p-1}~~\sum\limits_{j=(p^{r-2}+1)/2}^{p^{r-2}-1}d_{ip^{r-2}+j}^{(p^{r})}(X).
\end{equation}

We first prove some lemmas.

\begin{lemma}\label{Dmodp}
For $r\geq 2$ and $0\leq  l < p^{r}$, we have
$$
\left\{u \pmod {p^{r}} : u\in D_l^{(p^{r+1})} \right\} = D_{l \pmod
{p^{r-1}}}^{(p^{r})}
$$
and
$$
\{u \pmod p : u\in D_l^{(p^{r})}  \}  =\{1,2,\ldots, p-1\}.
$$
\end{lemma}
Proof. See \cite[Lemma 2]{DCH}. \qed

\begin{lemma}\label{card-VD}
For $r\geq 2$, let $v\in \mathbb{Z}^*_{p^{r}}$ and
$\mathcal{V}_v=\{v, v+p^{r},v+2p^{r}, \ldots,
v+(p-1)p^{r}\}\subseteq \mathbb{Z}^*_{p^{r+1}}$. If the Euler
quotient $Q_r(v)=\ell$, then we have
$$
\left|\mathcal{V}_v \cap
D_{\ell+ip^{r-1}\pmod{p^r}}^{(p^{r+1})}\right|=1
$$
for all $i: 0\leq i<p$.
\end{lemma}
Proof. By Eq.(\ref{eq:add struct2}). \qed

\begin{corollary}\label{VD}
For $r\geq 3$, the sum $\sum_{i=0}^{p-1} d_{\ell+ip^{r-1}\pmod{p^r}}(X)$ is divided
by $1+X^{p^{r}}+\dots+X^{(p-1)p^{r}}$. \end{corollary}

\begin{lemma}\label{card-CC}
Let $\mathcal{C}_0$ and $\mathcal{C}_1$ be defined with
$\mathfrak{r}\geq 3$ as above. Let $v\in
\mathbb{Z}^*_{p^{\mathfrak{r}-1}}$ and $\mathcal{V}_v=\{v,
v+p^{\mathfrak{r}-1},v+2p^{\mathfrak{r}-1}, \ldots,
v+(p-1)p^{\mathfrak{r}-1}\}\subseteq
\mathbb{Z}^*_{p^{\mathfrak{r}}}$.

 (1). If the Euler quotient $Q_{\mathfrak{r}-1}(v)=j+ip^{\mathfrak{r}-2}$ for some $0\leq i<p$ and $0\leq j\leq \frac{p^{\mathfrak{r}-2}-1}{2}$, we have
$$
|\mathcal{V}_v \cap \mathcal{C}_0|=(p+1)/2, ~~ |\mathcal{V}_v \cap
\mathcal{C}_1|=(p-1)/2.
$$

(2).  If the Euler quotient
$Q_{\mathfrak{r}-1}(v)=j+ip^{\mathfrak{r}-2}$  for some $0\leq i<p$
and $\frac{p^{\mathfrak{r}-2}+1}{2}\leq j\leq p^{\mathfrak{r}-2}-1$,
we have
$$
|\mathcal{V}_v \cap \mathcal{C}_0|=(p-1)/2, ~~ |\mathcal{V}_v \cap
\mathcal{C}_1|=(p+1)/2.
$$
\end{lemma}
Proof. By Lemma \ref{card-VD} and the definitions of $\mathcal{C}_0$
and $\mathcal{C}_1$. \qed

\begin{lemma}\label{poly-SS}
For $r\geq 3$, let $S^{(p^{r})}(X)$ and $\overline{S}^{(p^{r})}(X)$
be defined in Eq.(\ref{gene-poly}). Then we have
$$
S^{(p^{r})}(X) \equiv  \left\{
\begin{array}{rl}
S^{(p^{r-1})}(X)&  if ~~ p\equiv 1 \pmod 4,\\
\overline{S}^{(p^{r-1})}(X)&  if~~ p\equiv 3 \pmod 4,
\end{array}
\right. \pmod{X^{p^{r-1}}-1}.
$$
\end{lemma}
Proof. By  Lemma \ref{Dmodp} and Eq.(\ref{SX}) we see that
$$S^{(p^{r})}(X)\equiv  \frac{p+1}{2}~S^{(p^{r-1})}(X)+\frac{p-1}{2}~\overline{S}^{(p^{r-1})}(X)
\pmod{X^{p^{r-1}}-1}.$$ \qed

\begin{lemma}\label{poly-SS-value}
For $r\geq 3$, let $S^{(p^{r})}(X)$ and $\overline{S}^{(p^{r})}(X)$
be defined in Eq.(\ref{gene-poly}). For $0\leq j\leq r$, let
$\theta_j\in \overline{\mathbb{F}}_{2}$ be any primitive $p^{j}$-th
root of unity. If 2 is a primitive root modulo $p^2$, then  we have
$$
S^{(p^{r})}(\theta_j)\left\{
\begin{array}{ll}
=0, & \mathrm{if} ~~~ j=0,\\
=(p^{r-1}-1)/2, & \mathrm{if} ~~~ j=1,\\
\neq 0, &  \mathrm{otherwise},
\end{array}
\right.
$$
and
$$
\overline{S}^{(p^{r})}(\theta_j)\left\{
\begin{array}{ll}
=0, & \mathrm{if} ~~~ j=0,\\
=(p^{r-1}+1)/2, & \mathrm{if} ~~~ j=1,\\
\neq 0, &  \mathrm{otherwise}.
\end{array}
\right.
$$
\end{lemma}
Proof. By Theorem \ref{LC} and related arguments. \qed

\subsection{Technique for the proof}\label{keypoint}
%\textbf{Point 1.}

In this subsection, we always suppose $r\geq 3$.  Let
$S^{(p^{r})}(X)$ be defined in Eq.(\ref{gene-poly}) with
$wt(S^{(p^{r})}(X))=(p-1)(p^{r-1}-1)/2$, here and hereafter $wt(-)$
means the number of non-zero coefficients of a polynomial. It is
well-known that  the $k$-error linear complexity of
$(s_n^{(\mathfrak{r})})$ is computed by the following formula
$$
LC^{\F_2}_k((s_n^{(\mathfrak{r})}))=\min\limits_{0\leq wt(e(X))\leq k}\left\{ p^{\mathfrak{r}}-\deg \left(\gcd
(X^{p^{\mathfrak{r}}}-1,S^{(p^{\mathfrak{r}})}(X) +e(X))\right)\right\},
$$
where $e(X)\in \mathbb{F}_2[X]$ is the generating polynomial of an \emph{error-sequence}\footnote{It means that, $e_n=1$ if $s_n^{(\mathfrak{r})}$ is changed
when computing the $k$-error linear complexity of $(s_n^{(\mathfrak{r})})$,  and otherwise $e_n=0$.} of the same period of $(s_n^{(\mathfrak{r})})$.

 Let
$$
\Phi^{(p^{j})}(X)=\frac{X^{p^{j}}-1}{X^{p^{j-1}}-1}=1+X^{p^{j-1}}+
X^{2p^{j-1}}+\ldots+X^{(p-1)p^{j-1}} \in \mathbb{F}_2[X], ~~~ j\geq
1.
$$
We see that, if $2$ is a primitive root modulo $p^2$, each
$\Phi^{(p^{j})}(X)$ is irreducible and exactly has $(p-1)p^{j-1}$
many primitive $p^{j}$-th roots of unity in
$\overline{\mathbb{F}}_2$.

Let $e(X)=e_0+e_1X+\ldots+e_{p^{r}-1}X^{p^{r}-1}$ be a polynomial
over $\mathbb{F}_2$. We want to find an
$e(X)$ with the smallest $wt(e(X))$ such that
\begin{equation}\label{Hk}
\Phi^{(p^{r})}(X)|(S^{(p^{r})}(X)+e(X)).
\end{equation}
We suppose
\begin{equation}\label{pi}
S^{(p^{r})}(X)+e(X)= \Phi^{(p^{r})}(X)\cdot \pi^{(p^{r})}(X),
\end{equation}
where $\pi^{(p^{r})}(X)\in \mathbb{F}_2[X]$. Since
$\deg(S^{(p^{r})}(X)+e(X))<p^{r}$, we have
$\deg(\pi^{(p^{r})}(X))<p^{r-1}$ and
 $\pi^{(p^{r})}(X)$ should be one of the following:
$$
\begin{array}{l}
\pi^{(p^{r})}(X)=1;\\
\pi^{(p^{r})}(X)=X^{v_1}+X^{v_2}+\ldots+X^{v_{t}};\\
\pi^{(p^{r})}(X)=1+X^{v_1}+X^{v_2}+\ldots+X^{v_{t}};
\end{array}
$$
where $1\le t< p^{r-1}$ and $1\le v_1<v_2<\ldots<v_{t}<p^{r-1}$.

(i). If we use $\pi^{(p^{r})}(X)=1$, then $e(X)$ should be of the
form by (\ref{Hk})-(\ref{pi})
$$
e(X)=S^{(p^{r})}(X)+ \sum\limits_{0\leq j<p} X^{jp^{r-1}},
$$
which implies that
$wt(e(X))=wt(S^{(p^{r})}(X))+p=p^{r-1}(p-1)/2+(p+1)/2$.

(ii). If we use $\pi^{(p^{r})}(X)=X^{v_1}+X^{v_2}+\ldots+X^{v_{t}}$,
we let $\mathcal{I}=\{v_1, v_2,\ldots, v_{t}\}$ and
$\mathcal{V}_v=\{v, v+p^{r-1},v+2p^{r-1}, \ldots, v+(p-1)p^{r-1}\}$.
We also let $\mathcal{M}$ be the set of $v\in\mathbb{Z}^*_{p^{r-1}}$
such that $Q_{r-1}(v)=ip^{r-2}+j$ for $0\leq i<p$ and $0\leq j\leq
\frac{p^{r-2}-1}{2}$, and
$\mathcal{N}=\mathbb{Z}^*_{p^{r-1}}\setminus\mathcal{M}$. Then by
Eq.(\ref{pi}) we  derive that
$$
\begin{array}{rl}
e(X) = & \sum\limits_{v\in \mathcal{I}\cap
\mathcal{M}}~~\sum\limits_{n\in \mathcal{V}_v\cap
\mathcal{C}_0}X^{n} + \sum\limits_{v\in \mathcal{M}\setminus
\mathcal{I}}~~\sum\limits_{n\in \mathcal{V}_v\cap
\mathcal{C}_1}X^{n}
 \\
& + \sum\limits_{v\in \mathcal{I}\cap
\mathcal{N}}~~\sum\limits_{n\in \mathcal{V}_v\cap
\mathcal{C}_0}X^{n} + \sum\limits_{v\in \mathcal{N}\setminus
\mathcal{I}}~~\sum\limits_{n\in \mathcal{V}_v\cap
\mathcal{C}_1}X^{n} +\sum\limits_{v\in \mathcal{I}\setminus
\mathbb{Z}^*_{p^{r-1}}}~~\sum\limits_{0\le j<p}X^{v+jp^{r-1}},
\end{array}
$$
which implies by Lemma \ref{card-CC} that $wt(e(X))\geq
p^{r-2}(p-1)^2/2$ and the equality holds only if
\begin{equation}\label{eX}
e(X)=\sum\limits_{i=(p+1)/2}^{p-1}~~\sum\limits_{j=0}^{(p^{r-2}-1)/2}d_{ip^{r-2}+j}^{(p^{r})}(X)+
\sum\limits_{i=0}^{(p-3)/2}~~\sum\limits_{j=(p^{r-2}+1)/2}^{p^{r-2}-1}d_{ip^{r-2}+j}^{(p^{r})}(X).
\end{equation}
In this case (of equality holding),  we have
 $\Phi^{(p^{r})}(X)\mid (S^{(p^{r})}(X)+e(X))$ by Corollary
\ref{VD}.

 (iii). For
$\pi^{(p^{r})}(X)=1+X^{v_1}+X^{v_2}+\ldots+X^{v_{t}}$, we can get
similarly as (ii)
$$
\begin{array}{rl}
e(X) = & \sum\limits_{v\in \mathcal{I}\cap
\mathcal{M}}~~\sum\limits_{n\in \mathcal{V}_v\cap
\mathcal{C}_0}X^{n} + \sum\limits_{v\in \mathcal{M}\setminus
\mathcal{I}}~~\sum\limits_{n\in \mathcal{V}_v\cap
\mathcal{C}_1}X^{n}
 \\
& + \sum\limits_{v\in \mathcal{I}\cap
\mathcal{N}}~~\sum\limits_{n\in \mathcal{V}_v\cap
\mathcal{C}_0}X^{n} + \sum\limits_{v\in \mathcal{N}\setminus
\mathcal{I}}~~\sum\limits_{n\in \mathcal{V}_v\cap
\mathcal{C}_1}X^{n} +\sum\limits_{v\in \mathcal{I}\cup\{0\}\setminus
\mathbb{Z}^*_{p^{r-1}}}~~\sum\limits_{0\le j<p}X^{v+jp^{r-1}},
\end{array}
$$
which implies that $wt(e(X))\geq  p^{r-2}(p-1)^2/2+p$ and the
equality holds only if
$$
\begin{array}{rl}
e(X) = & \sum\limits_{i=(p+1)/2}^{p-1}~~\sum\limits_{j=0}^{(p^{r-2}-1)/2}d_{ip^{r-2}+j}^{(p^{r})}(X)\\
&+
\sum\limits_{i=0}^{(p-3)/2}~~\sum\limits_{j=(p^{r-2}+1)/2}^{p^{r-2}-1}d_{ip^{r-2}+j}^{(p^{r})}(X)
  +\sum\limits_{0\le j<p}X^{jp^{r-1}}.
\end{array}
$$

Therefore, from (i)-(iii) above, we conclude that the $e(X)$ in
(\ref{eX}) with the smallest weight $ p^{r-2}(p-1)^2/2$ can
guarantee $\Phi^{(p^{r})}(X)|(S^{(p^{r})}(X)+e(X))$. In other words,
for any $e(X)$ with $wt(e(X)))< p^{r-2}(p-1)^2/2$, we always have
$(S^{(p^{r})}(X)+e(X))\big|_{X=\theta_r}\neq 0$ for any primitive $p^{r}$-th root of unity $\theta_r\in \overline{\mathbb{F}}_{2}$ for $r\geq 3$.\\

%\textbf{Point 2}.

%If we replace $S^{(p^{r})}(X)$ by $\overline{S}^{(p^{r})}(X)$ in
%Point 1, the way in Point 1 helps us to get the smallest $wt(e(X))=
%p^{r-2}(p-1)^2/2$ such that
%$\Phi^{(p^{r})}(X)|(\overline{S}^{(p^{r})}(X)+e(X))$, in which case
%$e(X)$ is as in Eq.(\ref{eX}).

\subsection{Proof of the main theorem}

\textbf{Proof of Theorem 3}. First, from Subsection \ref{keypoint} we
find that $e(X)$ in (\ref{eX}) with $r=\mathfrak{r}\ge 3$ can guarantee
$\Phi^{(p^{\mathfrak{r}})}(X)\mid (S^{(p^{\mathfrak{r}})}(X)+e(X))$.
And hence by Eqs.(\ref{SX}) and (\ref{eX}) we have
\begin{equation}\label{SE-3}
S^{(p^{\mathfrak{r}})}(X)+e(X)=\sum\limits_{i=0}^{p-1}~~\sum\limits_{j=(p^{\mathfrak{r}-2}+1)/2}^{p^{\mathfrak{r}-2}-1}d_{ip^{\mathfrak{r}-2}+j}^{(p^{\mathfrak{r}})}(X)
\end{equation}
and $wt(e(X))=p^{\mathfrak{r}-2}(p-1)^2/2$ which is the smallest for
the assumption. From Eq.(\ref{SE-3}) we derive farther
$$
S^{(p^{\mathfrak{r}})}(X)+e(X)\equiv\sum\limits_{j=(p^{\mathfrak{r}-2}+1)/2}^{p^{\mathfrak{r}-2}-1}d_{j}^{(p^{\mathfrak{r}-1})}(X)=
S^{(p^{\mathfrak{r}-1})}(X) \pmod{X^{p^{\mathfrak{r}-1}}-1}.
$$
While by Lemma \ref{poly-SS-value}, we see that
$\Phi^{(p^{j})}(X)\nmid S^{(p^{\mathfrak{r}-1})}(X)$
for $j=2,3,\ldots,\mathfrak{r}-1$ and
$(X^p-1)|S^{(p^{\mathfrak{r}-1})}(X)$ if and only if $ p\equiv 1
\pmod 4$ or $ p\equiv 3 \pmod 4$ and $\mathfrak{r}$ is even.

This
means that
$$
\Phi^{(p^{j})}(X)\nmid
(S^{(p^{\mathfrak{r}})}(X)+e(X))
$$
for $2\leq j \leq \mathfrak{r}-1$, but
$$
\Phi^{(p^{\mathfrak{r}})}(X)(X^p-1)\mid
(S^{(p^{\mathfrak{r}})}(X)+e(X))
$$
for $p\equiv 1 \pmod 4$ or $p\equiv 3 \pmod 4$ and even $\mathfrak{r}>3$, or
$$
\Phi^{(p^{\mathfrak{r}})}(X)(X-1) \mid
(S^{(p^{\mathfrak{r}})}(X)+e(X))
$$
for $p\equiv 3 \pmod 4$ and odd $\mathfrak{r}\geq 3$.

 So we
get
$$
LC^{\F_2}_{p^{\mathfrak{r}-2}(p-1)^2/2}((s_n^{({\mathfrak{r}})}))=p^{\mathfrak{r}-1}-p
$$
if $ p\equiv 1 \pmod 4$ or $ p\equiv 3 \pmod 4$ and $\mathfrak{r}\geq 3$
is even, and otherwise
$$
LC^{\F_2}_{p^{\mathfrak{r}-2}(p-1)^2/2}((s_n^{({\mathfrak{r}})}))=p^{\mathfrak{r}-1}-1.
$$

Second, we consider $k<p^{\mathfrak{r}-2}(p-1)^2/2$. We note that in
this case, $\Phi^{(p^{\mathfrak{r}})}(X)\nmid
(S^{(p^{\mathfrak{r}})}(X)+e(X))$ for any $e(X)$ with
$wt(e(X))=k<p^{\mathfrak{r}-2}(p-1)^2/2$. By Lemma \ref{poly-SS}, it is
reduced to consider $LC^{\F_2}_k((s_n^{({\mathfrak{r}-1})}))$ if
$p\equiv 1 \pmod 4$, or
$LC^{\F_2}_k((\overline{s}_n^{({\mathfrak{r}-1})}))$ if $p\equiv 3
\pmod 4$. That is,
$$
LC^{\F_2}_{k}((s_n^{({\mathfrak{r}})}))=(p^{\mathfrak{r}}-p^{\mathfrak{r}-1})+\left\{
\begin{array}{ll}
LC^{\F_2}_{k}((s_n^{({\mathfrak{r}-1})})), & \mathrm{if} ~~~ p\equiv 1 \pmod 4,\\
LC^{\F_2}_{k}((\overline{s}_n^{({\mathfrak{r}-1})})), & \mathrm{if}
~~~ p\equiv 3 \pmod 4.
\end{array}
\right.
$$
To conclude the proof, it remains to note that
$LC^{\F_2}_k((s_n^{(\mathfrak{r}-1)}))=0$ for $k\ge
(p^{\mathfrak{r}-2}-1)(p-1)/2$ and
$LC^{\F_2}_{k}((\overline{s}_n^{({\mathfrak{r}-1})}))=0$ for $k\ge
(p^{\mathfrak{r}-2}+1)(p-1)/2$. \qed

\subsection{Further discussions}

Theorem \ref{klc-r} gives us a recurrent formula for the calculation of the
$k$-error linear complexity of $(s_n^{(\mathfrak{r})})$. It relates to $((\overline{s}_n^{(\mathfrak{r})}))$ when $p\equiv 3 \pmod 4$.
So here we give a result of  $LC^{\F_2}_k((\overline{s}_n^{(\mathfrak{r})}))$ for $\mathfrak{r}=2$.

\begin{theorem}\label{klc-s-bar-2-primitive}
Let $(\overline{s}_n^{(\mathfrak{r})})$ be the binary sequence over
$\mathbb{F}_2$ defined in Eq.(\ref{euler-binary-complement}).
 If $2$ is a primitive root modulo $p^2$, then
the $k$-error linear complexity  of  $(\overline{s}_n^{(\mathfrak{r})})$ for $\mathfrak{r}=2$ satisfies
\[
LC^{\F_2}_k((\overline{s}_n^{(2)}))=\left\{
\begin{array}{cl}
p^2-1, & \mathrm{if}\,\ k=0, \\
p^2-p+1, & \mathrm{if}\,\ 0< k<p-1, \\
p^2-p, & \mathrm{if}\,\ p-1\le k<(p-1)^2/2, \\
p-1, & \mathrm{if}\,\  (p-1)^2/2\le k<(p^2-1)/2, \\
0, & \mathrm{if}\,\ k\ge (p^2-1)/2,
\end{array}
\right.\\
\]
 if $p \equiv 1 \pmod 4$, and otherwise
\[
 LC^{\F_2}_k(\overline{s}_n^{(2)})=\left\{
\begin{array}{cl}
p^2-p, & \mathrm{if}\,\ 0\le k< (p-1)^2/2, \\
p-1, & \mathrm{if}\,\  (p-1)^2/2 \le k < (p^2-1)/2, \\
0, & \mathrm{if}\,\ k\ge  (p^2-1)/2.
\end{array}
\right.
\]
\end{theorem}

The proof (of Theorem \ref{klc-s-bar-2-primitive}) can be done following the way in Subsection 2.2 or
\cite[pp.6-7]{CNW}, so we omit it.\\

 Another thing we need to say is that, due to the fact that  the recurrent formula in Theorem \ref{klc-r} does not cover all $k$,
 we cannot determine the exact values of $LC^{\F_2}_k(s_n^{(\mathfrak{r})})$
 for $p^{\mathfrak{r}-3}(p-1)^2/2<k <(p^{\mathfrak{r}-2}-1)(p-1)/2$ when $\mathfrak{r}\geq 4$.
 But of course, we have a bound on it for $p^{\mathfrak{r}-3}(p-1)^2/2<k <(p^{\mathfrak{r}-2}-1)(p-1)/2$
 $$
 p^{\mathfrak{r}}-p^{\mathfrak{r}-1} \leq LC^{\F_2}_k(s_n^{(\mathfrak{r})}) \leq p^{\mathfrak{r}}-1.
 $$
However, we have exact formula when $\mathfrak{r}=3$ and we state it as follows

\begin{corollary}\label{klc-r=3}
Let $(s_n^{(\mathfrak{r})})$ be the binary sequence over
$\mathbb{F}_2$ defined in Eq.(\ref{binarythreshold}) or
Eq.(\ref{euler-binary}).
 If $2$ is a primitive root modulo $p^2$, then
the $k$-error linear complexity  of  $(s_n^{(\mathfrak{r})})$ for $\mathfrak{r}=3$
satisfies
\[
 LC^{\F_2}_k((s_n^{(3)}))=\left\{
\begin{array}{cl}
p^3-p, & \mathrm{if}\,\ 0\le k<(p-1)^2/2, \\
p^3-p^2, & \mathrm{if}\,\ (p-1)^2/2\leq k<p(p-1)^2/2,\\
p^2-p, & \mathrm{if}\,\ k=p(p-1)^2/2,
\end{array}
\right.
\]
 if $p \equiv 1 \pmod 4$, and otherwise
 \[
 LC^{\F_2}_k((s_n^{(3)}))=\left\{
\begin{array}{cl}
p^3-p, & \mathrm{if}\,\ 0\le k<(p-1)^2/2, \\
p^3-p^2+p-1, & \mathrm{if}\,\ (p-1)^2/2\leq k<(p^2-1)/2,\\
p^3-p^2, & \mathrm{if}\,\ (p^2-1)/2\leq k<p(p-1)^2/2,\\
p^2-1, & \mathrm{if}\,\ k=p(p-1)^2/2.
\end{array}
\right.\\
\]
\end{corollary}

We also run a program to confirm our results for some examples.

(1). Let $p=3$ and $\mathfrak{r}=3$. We choose $g=11$ then we have
$$D_5^{(p^{3})}=\{4,23\}, D_6^{(p^{3})}=\{10,17\},
D_7^{(p^{3})}=\{2,25\}, D_8^{(p^{3})}=\{5,22\}.$$
And we get
 \[
 LC^{\F_2}_k((s_n^{(3)}))=\left\{
\begin{array}{cl}
24, & \mathrm{if}\,\ 0\leq k \leq 1, \\
 20, & \mathrm{if}\,\ 2\leq k\leq 3,\\
18, & \mathrm{if}\,\ 4\leq k\leq 5,\\
8, & \mathrm{if}\,\  k=6,
\end{array}
\right.
\]
which coincides with Corollary \ref{klc-r=3}.\\

(2). Let $p=5$ and $\mathfrak{r}=3$. We choose $g=3$ then we have
 $$
  D_{13}^{(p^{3})}=\{73,89,52,36\}, D_{14}^{(p^{3})}=\{94,17,31,108\}, D_{15}^{(p^{3})}=\{32,51,93,74\},
 $$
 $$
 D_{16}^{(p^{3})}=\{96,28,29,97\},
 D_{17}^{(p^{3})}=\{38,84,87,41\},
 D_{18}^{(p^{3})}=\{114,2,11,123\},
 $$
 $$
 D_{19}^{(p^{3})}=\{92,6,33,119\},
 D_{20}^{(p^{3})}=\{26,18,99,107\},
 D_{21}^{(p^{3})}=\{78,54,47,71\},
 $$
 $$
 D_{22}^{(p^{3})}=\{109,37,16,88\},
 D_{23}^{(p^{3})}=\{77,111,48,14\},
 D_{24}^{(p^{3})}=\{106,83,19,42\}.
 $$
And we get
 \[
 LC^{\F_2}_k((s_n^{(3)}))=\left\{
\begin{array}{cl}
120, & \mathrm{if}\,\ 0\leq k \leq 7, \\
100, & \mathrm{if}\,\ 8\leq k\leq 39,\\
20, & \mathrm{if}\,\ k=40,
\end{array}
\right.
\]
which coincides with Corollary \ref{klc-r=3}.

\section{Final remarks and conclusions}

In  this work, a progress is made to determine the $k$-error linear
complexity of a family of binary sequences derived from Euler
quotients modulo $p^{\mathfrak{r}}$.

It is interesting to consider the $k$-error linear
complexity of
these binary sequences over $\mathbb{F}_p$. Such kind of work has been done before, see e.g. \cite{AW}.
From the proof of Lemma \ref{poly-SS}, we also can get a recurrent formula on $ LC^{\mathbb{F}_p}_k((s_n^{(\mathfrak{r})}))$ as in Theorem \ref{klc-r}.
Some techniques in \cite{CNW} can be helpful for this problem.

We finally remark that, very recently a new family of binary
sequences, introduced by Xiao, Zeng, Li and Helleseth \cite{XZLH},
are in fact related to  the Fermat-Euler quotients. Let $p-1=ef$ and
$g$ a primitive root modulo $p^2$ as in Sect.2. Since
$$
\mathbb{Z}_{p^{\mathfrak{r}}}=\mathbb{Z}_{p^{\mathfrak{r}}}^* \cup
p\mathbb{Z}_{p^{\mathfrak{r}-1}}^*\cup
p^2\mathbb{Z}_{p^{\mathfrak{r}-2}}^* \cup\cdots \cup
p^{\mathfrak{r}-1}\mathbb{Z}_{p}^*\cup \{0\},
$$
we define generalized cyclotomic classes for each
$\mathbb{Z}_{p^r}^*$, where $1\leq r\leq \mathfrak{r}$, in the
following way
$$
D_0^{(p^r,f)}\triangleq \langle g^{fp^{r-1}} \rangle=\{g^{kfp^{r-1}}
\pmod {p^r} : 0\leq k<e\}
$$
and
$$
D_l^{(p^r,f)}\triangleq g^{l} D_0^{(p^r,f)}=\{g^l\cdot g^{kfp^{r-1}}
\pmod {p^r} : 0\leq k<e \}, ~~ 1\leq l<fp^{r-1}.
$$
Indeed $D_0^{(p^r,f)}, D_1^{(p^r,f)},\ldots,
D_{fp^{r-1}-1}^{(p^r,f)}$ give a partition of $\mathbb{Z}_{p^r}^*$.

We find that when $\mathfrak{r}\geq 2$,
$$
\bigcup\limits_{i=0}^{f-1} D_{l+ip^{\mathfrak{r}-1}}^{(p^{\mathfrak{r}},f)}=D_{l}^{(p^{\mathfrak{r}})}, ~~~ 0\leq
l<p^{\mathfrak{r}-1}
$$
where $D_{l}^{(p^{\mathfrak{r}})}$ is defined by Euler quotient in
Sect.2.

Xiao, Zeng, Li and Helleseth \cite{XZLH} considered the following
binary sequences
$$
t_n^{(\mathfrak{r})}=\left\{
\begin{array}{ll}
0, & \mathrm{if} ~ n\pmod {p^{\mathfrak{r}}} \in \mathcal{C}_0,\\
1, &  \mathrm{if} ~  n\pmod {p^{\mathfrak{r}}} \in \mathcal{C}_1,
\end{array}
\right.
$$
where
$$
\mathcal{C}_0=\bigcup_{r=1}^{\mathfrak{r}}~~~\bigcup_{l=p^{r-1}f/2}^{p^{r-1}f-1}p^{\mathfrak{r}-r}D_{l+b
\pmod {p^{r-1}f}}^{(p^r,f)},
$$
$$
\mathcal{C}_1=\bigcup_{r=1}^{\mathfrak{r}}~~~\bigcup_{l=0}^{p^{r-1}f/2-1}p^{\mathfrak{r}-r}D_{l+b
\pmod {p^{r-1}f}}^{(p^r,f)}\cup \{0\}
$$
for $b\in \mathbb{Z} : 0\leq b<p^{\mathfrak{r}-1}f$ and even $f$.

When $\mathfrak{r}=2$, Xiao et al \cite{XZLH} proved the linear
complexity of $(t_n^{(\mathfrak{r})})$ for $f=2^r$ and we (joint
with other coauthor) studied the $k$-error linear complexity of
$(t_n^{(\mathfrak{r})})$ for $f=2$ in \cite{WXCK}. We note here that the result in \cite{WXCK}
can be extended to the case of $f=2^r$ for any $r\geq 2$.

When $\mathfrak{r}\geq 3$, the second author of this article (partly
joint with other coauthors) proved the linear complexity of
$(t_n^{(\mathfrak{r})})$ for $f=2^r$, which gave a positive answer
to a conjecture stated in \cite{XZLH}, see \cite{E}.  All known
results indicate that $(t_n^{(\mathfrak{r})})$ has nice
cryptographic features.
Following the way of \cite{WXCK} and this work, one can discuss the
$k$-error linear complexity of $(t_n^{(\mathfrak{r})})$ for
$\mathfrak{r}\geq 3$.

\section*{Acknowledgements}

The work was partially supported by the National Natural Science
Foundation of China under grant No.~61772292.


\begin{thebibliography}{99}


\bibitem{ADS} T. Agoh, K. Dilcher and L. Skula.
Fermat quotients for composite moduli.  J. Number Theory 66 (1997) 29-50.

\bibitem{AW} H. Aly, A. Winterhof.
On the k-error linear complexity over $\mathbb{F}_p$ of Legendre and Sidelnikov sequences. Des. Codes Cryptography 40(3) (2006) 369-374.

%\bibitem{AS}A. Akbary and S. Siavashi. The Largest Known Wieferich Numbers. Integers 18(\#A3) : 1-6 (2018).



\bibitem{BFKS}
J. Bourgain, K. Ford, S. Konyagin and I. E. Shparlinski. On the
divisibility of Fermat quotients. Michigan Math. J. 59 (2010) 313-328.


\bibitem{Chang}
M. C. Chang. Short character sums with Fermat quotients. Acta Arith.
152 (2012) 23-38.



\bibitem{C} Z. Chen. Trace representation and linear complexity of binary sequences derived from Fermat quotients. Sci. China  Inf. Sci. 57 (11) (2014) 11:2109.

\bibitem{CD}  Z. Chen,  X. Du. On the linear complexity of binary threshold sequences derived from
Fermat quotients. Des. Codes Cryptogr. 67 (2013) 317--323.

\bibitem{CDR2015} Z. Chen,  X. Du, R. Marzouk.  Trace representation of pseudorandom binary
sequences derived from Euler quotients. Appl. Algebra Eng. Commun. Comput. 26(6) (2015) 555-570.

\bibitem{CNW} Z. Chen, Z. Niu, C. Wu. On the $k$-error linear complexity of binary
sequences derived from polynomial quotients. Sci. China  Inf. Sci. 58 (09) (2015) 09:2107.



\bibitem{CW2} Z. Chen  and A.  Winterhof. Additive character sums of
polynomial quotients. Theory and Applications of Finite Fields-Fq10,
pp.67-73, Contemp. Math., 579, Amer. Math. Soc., Providence, RI, 2012.



\bibitem{CW3} Z. Chen and A. Winterhof. Interpolation of Fermat quotients. SIAM J. Discr. Math. 28 (2014) 1-7.


\bibitem{COW}  Z. Chen, A. Ostafe, A. Winterhof. Structure of
pseudorandom numbers derived from Fermat quotients. In: Proceedings
of the 3rd International Conference on Arithmetic of Finite Fields-WAIFI2010, pp.73-85, Lecture Notes in Comput Sci, vol. 6087, Berlin: Springer-Verlag, 2010.


\bibitem{CDR}
 T. W. Cusick, C. Ding and A. Renvall. Stream Ciphers and Number Theory. Gulf Professional Publishing, 2004.




\bibitem{DXS}
C. Ding, G. Xiao and W. Shan. The Stability Theory of Stream Ciphers. Lecture
Notes in Comput. Sci. vol. 561. Springer-Verlag, Berlin, 1991.

\bibitem{DCH}
 X. Du, Z. Chen, L. Hu. Linear complexity of binary
sequences derived from Euler quotients with prime-power modulus.
Inform. Process. Lett. 112 (2012) 604-609.


\bibitem{E}
V. Edemskiy. The linear complexity of new binary cyclotomic sequences of period $p^n$. CoRR abs/1712.03947 (2017)

\bibitem{GW}  D. G\'{o}mez-P\'{e}rez, A. Winterhof. Multiplicative character sums of
Fermat quotients and pseudorandom sequences. Period. Math. Hungar. 64 (2012) 161-168.

%\bibitem{Massey} J. Massey. Shift register synthesis and BCH decoding. IEEE Trans Inform Theory  15 (1969) 122--127.

\bibitem{NCD}
Z. Niu, Z. Chen, X. Du:
Linear complexity problems of level sequences of Euler quotients and their related binary sequences. Sci. China  Inf. Sci. 59(3) (2016) 3:2106.



\bibitem{OS} A. Ostafe  and  I. E. Shparlinski. Pseudorandomness and dynamics of
Fermat quotients. SIAM J. Discr. Math. 25 (2011) 50-71.

\bibitem{Sha}
M. Sha. The arithmetic of Carmichael quotients. Period. Math. Hungar. 71(1) (2015) 11-23.

\bibitem{S}I. E. Shparlinski. Character sums with Fermat quotients. Quart. J. Math. 62(4) (2011) 1031-1043.

\bibitem{S2010} I. E. Shparlinski.  Bounds of multiplicative character sums with
Fermat quotients of primes. Bull. Aust. Math. Soc. 83(3) (2011)
456-462.

\bibitem{S2011} I. E. Shparlinski.  On the value set of Fermat
quotients. Proc. Amer. Math. Soc. 140(4) (2012) 1199-1206.

\bibitem{S2011b} I. E. Shparlinski.  Fermat quotients: Exponential sums, value set and primitive
roots. Bull. Lond. Math. Soc.  43(6) (2011) 1228-1238.

\bibitem{SW}
I. E. Shparlinski  and A. Winterhof. Distribution of values of
polynomial Fermat quotients. Finite Fields Appl. 19 (2013)
93-104.




\bibitem{SM}
M. Stamp and C. F. Martin. An algorithm for the $k$-error linear complexity of
binary sequences with period $2^n$. IEEE Trans. Inf. Theory 39 (1993) 1398-1401.





\bibitem{WXCK}
C. Wu, C. Xu, Z. Chen, P. Ke.
On error linear complexity of new generalized cyclotomic binary sequences of period $p^2$. CoRR abs/1711.06063 (2017)


\bibitem{XZLH} Z. Xiao, X. Zeng, C. Li, T. Helleseth.
New generalized cyclotomic binary sequences of period $p^2$. Des. Codes Cryptogr. https://doi.org/10.1007/s10623-017-0408-7.(2017)



\end{thebibliography}
\end{document}